\documentclass[submission,copyright,creativecommons]{eptcs}
\providecommand{\event}{UNIF 2010} 
\usepackage{breakurl}             

\usepackage{amsmath} \usepackage{amsthm} \usepackage{amscd} 
\usepackage{amssymb} \usepackage{amsfonts} \usepackage{array}
\usepackage{graphics} \usepackage[all]{xy} \usepackage{color}

\newtheorem{definition}{Definition}

\newtheorem{theorem}{Theorem}

\newcommand{\msf}[1]{{\mathsf{#1}}}

\newcommand{\mi}[1]{{\mathit{#1}}}

\newcommand{\std}{\mathsf{STD}}
\newcommand{\pk}{\mi{pk}}
\newcommand{\sh}{\mi{sh}}
\newcommand{\acun}{\mathsf{ACUN}}
\newcommand{\StdOps}{\mi{StdOps}}
\newcommand{\lra}{\leftrightarrow}
\newcommand{\Lra}{\Leftrightarrow}

\newcommand{\Ra}{\Rightarrow}

\newcommand{\Vars}{\mi{Vars}}

\newcommand{\txor}{\texttt{XOR}}
\newcommand{\xorseq}[3]{#1\oplus#2 \oplus #3}
\newcommand{\mxor}[2]{#1\oplus #2}

\newcommand{\Terms}{\mi{Terms}}

\newcommand{\subterm}{\mi{subterm}}

\newcommand{\fa}{\forall}
\newcommand{\ex}{\exists}

\newcommand{\SubTerms}{\mi{SubTerms}}
\newcommand{\op}[1]{\mi{op_{#1}}}
\newcommand{\napprox}{\not\approx}

\newcommand{\lan}{\langle}
\newcommand{\ran}{\rangle}
\newcommand{\sqss}{\sqsubset}
\newcommand{\Constants}{\mi{Constants}}

\newcommand{\tacun}{\textsf{ACUN}}

\newcommand{\VarIdP}{\mi{VarIdP}}

\newcommand{\pure}{\msf{pure}}
\newcommand{\tuple}[2]{\lan #1, ~#2 \ran}

\newcommand{\Operators}{\mi{Operators}}

\newcommand{\Ops}{\mi{Ops}}

\newcommand{\penc}[2]{[#1]^{\to}_{#2}}
\newcommand{\senc}[2]{[#1]^{\lra}_{#2}}

\newcommand{\dnut}{\textsf{DNUT}}
\newcommand{\dnutsat}{\textsf{DNUT}-Satisfying}

\newcommand{\Th}{\mi{Th}}
\newcommand{\eqop}{\mi{eqop}}
\newcommand{\interm}{\mi{interm}}
\newcommand{\eqth}{\msf{EQTH}}
\newcommand{\feqop}{\msf{FEQOP}}

\title{Disabling equational theories in unification for cryptographic protocol analysis through tagging}
\author{Sreekanth Malladi
\institute{Dakota State University, \\
Madison, SD - 57042, USA}
\email{malladis@pluto.dsu.edu}
}

\def\titlerunning{Disabling equational theories in unification}
\def\authorrunning{Sreekanth Malladi}

\begin{document}

\maketitle

\begin{center} \today \end{center}

\section{Introduction}\label{s.intro}

Most of the research on protocol security in the past two decades has been conducted assuming a free message algebra. However, operators such as Exclusive-OR (\txor) possess algebraic properties. There were instances when a protocol was secure in the free algebra, but insecure in the presence of equational theories induced by such operators~\cite{RS98}. Hence, it is important to conduct protocol analysis with careful consideration of equational theories.

Unification is an important part of symbolic protocol analysis that is affected by equational theories. If we can disable them, i.e., if we can construct protocol messages such that unification in the presence of equational theories implies the same unification in their absence, then it is a good step in simplifying protocol analysis in the presence of theories.

This is the point we consider in this paper. We formulate a new tagging scheme for protocol messages that essentially disables disjoint equational theories\footnote{Disjoint theories are those where the equations in the theories do not share operators.}. As a consequence of this result, we could recently  achieve the following, immensely useful result for protocols involving the \txor{} operator that possesses the \tacun{} theory:

\begin{quote}

	Under a certain tagging scheme, if a protocol is secure under a free algebra, then it is secure in the presence of the {\sf ACUN} theory.

\end{quote}

We provide a full formal proof of this result in~\cite{Mall10C}. This result essentially disables the {\sf ACUN} theory from having any effect on protocol security. Further, it allows us to lift many existing results obtained under a free algebra. For instance, the classical ``small-system" decidability result by Lowe in his pioneer work~\cite{Lowe99} states,

\begin{quote}

``If there is no attack on a small system of a protocol (with exactly one agent playing each role of the protocol), leading to a breach of secrecy, then there is no attack on any larger system leading to a breach of secrecy."

\end{quote}

Although Lowe has achieved this result in a free term algebra, we can tag protocols in our scheme, and use our main result of \cite{Mall10C} to conclude that the small-system result is valid even under the {\sf ACUN} theory, since no new attacks are enabled. We can similarly recover many existing results achieved under a free algebra, such as simplifying transformations for protocols~\cite{HL99}, preventing type-flaw and multi-protocol attacks~\cite{HLS00,TG00}.

Such a similar result is possible under other theories such as $\msf{ACU} \cup \msf{Inverse}$ and $\msf{ACU} \cup \msf{Idempotence}$ as well. However, while a crucial component of~\cite{Mall10C} is disabling equational unification (which is the only  point of this paper), the protocol analysis framework used in~\cite{Mall10C} is from~\cite{Chev04}, which is tailored only to the \tacun{} theory. To achieve a similar result under other theories, we would have to use the result of the current paper in suitable protocol models such as \cite{DLLT08}. This is a topic of current research.

It is very important to note that our result does \underline{\em not} consider equations of the form $[a, b] \oplus [c, d] = [a\oplus c, b \oplus d]$, which would hold when the operator $\oplus$ is homomorphic. The reason is that, we use the algorithm by Baader \& Schulz for combined theory unification~\cite{BS96} to achieve our result. The algorithm cannot handle  such equations that use operators in disjoint theories  (the above equation uses pairing, which is a free operator and the \txor{} operator). However, some implementations could lead to such equations, and we consider it an important  direction of future research to include them.

\section{Term Algebra}\label{s.term-algebra}

We will assume the existence of a basic, indivisible set of terms called variables and constants denoted as {\em Vars} and {\em Constants} respectively. We define a set of operators, $\mi{Ops} = \StdOps \cup \eqop$, where, $\StdOps = \{ \mi{sequence}$, $\mi{penc}, \mi{senc}, \pk, \sh \}$.  We use some syntactic sugar in using some of these operators:\\
  $\mi{sequence}(t_1,\ldots,t_n) = [t_1, \ldots, t_n]$, 
	$\mi{penc}(t,k)       =   [t]^{\to}_k$, 
	$\mi{senc}(t,k) 			=   {[t]^{\lra}_k}$,  
	$\eqop(t_1,\ldots,t_n)		=		t_1 \oplus \ldots \oplus t_n$.

Note that, although we use the symbol $\oplus$  for $\eqop$ that is conventionally used for \txor{}, here we treat it as a general operator that has some equational theory.

The term algebra is the infinite set, {\em Terms}, where $\Vars \cup \Constants \subset \Terms$ and $(\fa t_1,\ldots,t_n \in \Terms; f \in \Ops)(f(t_1,\ldots,t_n) \in \Terms)$. We will define two relations, $\subterm{}$ and $\interm{}$ denoted $\sqsubset$ and $\Subset$ respectively on terms such that:

\begin{itemize}

	\item $t \sqss t'$ iff $t = t'$ or $t' = f(t_1,\ldots,t_n)$ where $f \in 						\Ops$ and $t \sqss t''$ for some $t'' \in \{ t_1, \ldots, t_n \}$.

	\item $t \Subset t'$ iff 
	$(\ex t_1,\ldots,t_m; i \in \{1,\ldots,m\})( (t_1 \oplus \ldots \oplus t_m = t') \wedge (t_i = t))$.	
	
	\item $\SubTerms(T) = \{ t \mid (\ex t' \in T)(t \sqsubset t') \}$.

\end{itemize}

Interms are also subterms, but subterms are not necessarily interms. For instance, $[1,a]$ is both an interm and a subterm of $[1,a] \oplus [2,b] \oplus [3, [n_b]_k]$, but $n_b$ is only a subterm in it, not an interm.

\begin{definition}{\em\bf [Equation and Theory]}\label{d.equation-theory}
An {\em equation} is a tuple $\tuple{\mi{term}}{\mi{term}}$. We write $t_1 =_{e} t_2$ if $e = \tuple{t_1}{t_2}$ is an equation. A {\em theory} is a set of equations. If $\Th$ is a theory, we write $t =_{\Th} t'$, if there exist a finite sequence of equations $e_1, \ldots, e_n \in \Th$ such that, $t =_{e_1} t_1$, $t_1 =_{e_2} t_2$, $\ldots$, $t_{n-1} =_{e_n} t'$.

\end{definition}

The theory $\std$ for $\StdOps$ is a set of equations between syntactically equal terms: 

$\{ \tuple{[t_1,\ldots,t_n]}{[t_1,\ldots,t_n]}$, $\tuple{\senc{t}{k}}{\senc{t}{k}}$, $\tuple{\penc{t}{k}}{\penc{t}{k}}$,  $\tuple{\pk(t)}{\pk(t)}$, $\tuple{\sh(t_1,t_2)}{\sh(t_1,t_2)} \}$.

The theory $\eqth$ has equations solely with the $\eqop$ ($\oplus$) operator. For our main result, we will consider the \tacun{} theory as $\eqth$, but in principle, this can be any set of equations where $\StdOps$ are not used. There can also be multiple operators in $\eqth$:
 
$\{ \tuple{ t_1 \oplus (t_2 \oplus t_3) }{ (t_1 \oplus t_2) \oplus t_3 }$,  $\tuple{\mxor{t_1}{t_2}}{\mxor{t_2}{t_1}}$, 
	$\tuple{\mxor{t}{0}}{t}$, $\tuple{\mxor{t}{t}}{0} \}$. 

We also define $\feqop$, which is a theory in which the $\oplus$ operator is free: 

 $\feqop =  \{ \tuple{t_1 \oplus \ldots \oplus t_n}{t_1 \oplus \ldots \oplus t_n} \}$.

\begin{definition}{\em\bf [Operators]}\label{d.operators}

	Let $\Operators(\Th)$ denote all the operators used to form the equations in the theory $\Th$\footnote{We use an underscore (\_) in a formula, when the value in it doesn't affect the truthness of the formula.}:
	
	\[
	\Operators(\Th) = 
	\left\{ 
	\begin{array}{ll}
		\op{} \mid 
				(\ex e \in \Th; t_1, t_2, t \in \Terms)
				\left(
					\begin{array}{c}
						((t \sqss t_1) \vee (t \sqss t_2)) \wedge \\
						(e = \tuple{t_1}{t_2})	\wedge 
						(t = \op{}(\_,\ldots,\_))					
					\end{array}				
				\right)
	\end{array}
	\right\}.
\]	

Theories $\Th_1$ and $\Th_2$ are {\em disjoint} if $\mi{Operators}(\Th_1) \cap \mi{Operators}(\Th_2) = \{\}$.

\end{definition}

We will say that a term $t$ is {\em pure} wrt the theory $\Th$, if all of its subterms are made only from $\mi{Operators}(\Th)$: $\pure(t,\Th) \Lra (\fa op(\_,\ldots,\_) \sqsubset t)(op \in \mi{Operators}(\Th)).$

We will now consider equational unification. We will abbreviate ``Unification Algorithm" to UA and ``Unification Problem" to UP:

\begin{definition}{\em\bf [Unification Problem, Unifier, Unification Algorithm]}\label{d.unification}

A {\em $\Th$-UP} is a tuple of terms $\tuple{m}{t}$ denoted $m \stackrel{?}{\approx}_{\Th} t$, where $m$ and $t$ are pure wrt $\Th$. If $\Th$ is a theory, a set of {\em $\Th$-UPs}, $\Gamma$, is {\em $\Th$-Unifiable} with a set of substitutions $\sigma$ called a {\em $\Th$-Unifier}, if  $(\fa m \stackrel{?}{\approx}_{\Th} t \in \Gamma)(m\sigma =_{\Th} t\sigma)$. A {\em $\Th$-Unifier} $\sigma$ is a {\em most general $\Th$-Unifier}, for a set of {\em $\Th$-UPs} $\Gamma$, if every other $\Th$-Unifier $\rho$ for $\Gamma$ is such that, $\rho = \sigma\rho$. 
A {\em complete $\Th$-UA} returns all possible most general $\Th$-Unifiers for any set of $\Th$-UPs.

\end{definition}

UAs for two disjoint theories $\Th_1$ and $\Th_2$, may be combined to output the unifiers for a set of $(\Th_1 \cup \Th_2)$-UPs using Baader \& Schulz Combination Algorithm (BSCA)~\cite{BS96}. We give a more detailed explanation in Appendix~\ref{s.BSCA}, using an example UP for the interested reader.

BSCA first takes as input, a set of $(\Th_1 \cup \Th_2)$-UPs, say $\Gamma$, and applies some transformations on them to derive $\Gamma_{5.1}$ and $\Gamma_{5.2}$ that are sets of $\Th_1$-UPs and $\Th_2$-UPs respectively. It then combines the unifiers for $\Gamma_{5.1}$ and $\Gamma_{5.2}$ obtained using $\Th_1$-UA and $\Th_2$-UA respectively (see Appendix~\ref{s.BSCA}, Def.~\ref{d.Combined-Unifier}) to form the unifier(s) for $\Gamma$.
Further, if all UPs in $\Gamma_{5.1}$ and $\Gamma_{5.2}$ are $\Th_1$-Unifiable and $\Th_2$-Unifiable respectively, then $\Gamma$ is $(\Th_1 \cup \Th_2)$-Unifiable. 

It has been proven in~\cite{BS96} that the combined unifier obtained is a complete $(\Th_1 \cup \Th_2)$-UA for any $(\Th_1 \cup \Th_2)$-UP if $\Th_1$-UA and $\Th_2$-UA are complete and if $\Th_1$ and  $\Th_2$ are disjoint.

\section{{\sf DNUT} - Disabling Non-Unifiability of Terms}\label{s.DNUT}

We now state our main requirement on terms, namely {\sf DNUT}. 

\begin{definition}[{\sf DNUT}]\label{d.DNUT}

A set of terms $T$ is {\em \dnutsat{}} or {\em \dnutsat}$(T)$ iff:

\begin{enumerate}

	\item No two interms of an $\eqop$ term are $\std$-{\sf Unifiable}\footnote{$\mathbb{N}$ is the set of natural numbers.}:
	
				\[	(\fa t \in \SubTerms(T); n \in \mathbb{N})
					\left(
						\begin{array}{c}
							(t = t_1 \oplus \ldots \oplus t_n) \wedge (n > 1) \wedge \\
							(\fa i, j \in \{1, \ldots, n\})((i \neq j) \Ra (t_i \napprox_{\std} t_j))								
						\end{array}
					\right).
				\]

	\item No interm of an $\eqop$ term is $\std$-{\sf Unifiable} with an interm of any other $\eqop$ term:

\[		(\fa t, t' \in \SubTerms(T))
					\left(
						(\ex t_1, t'_1)
							((t_1 \Subset t) \wedge 
							(t'_1 \Subset t')
						\Ra									
							(t_1 \napprox_{\std} t'_1))
					\right).
\]

\item The Unity element is not a part of any $\eqop$ term: 

\[	(\fa t_1 \oplus \ldots \oplus t_n \in \SubTerms(T); n \in \mathbb{N})\left( (\not\ex i \in \{1,\ldots,n\})(t_i = 0) \right).	\]

\end{enumerate}

\end{definition}

The first requirement of \dnut{} can be satisfied by ensuring that every term in the set $\{ t_1, \ldots, t_n \}$ is a pair that starts with a distinct constant, if $t_1 \oplus \ldots \oplus t_n$ is a subterm of $T$. For instance, consider, $A \oplus N_B \oplus \penc{N_A}{\pk(B)} \oplus \senc{N_B}{K}$. This can be changed to $[1, A] \oplus [2, N_B] \oplus [3, \penc{N_A}{\pk(B)}] \oplus [4,\senc{N_B}{K}]$ so that, no two terms in the set $\{ [1, A], [2, N_B], [3, \penc{N_A}{\pk(B)}], [4,\senc{N_B}{K}] \}$ are $\std$-Unifiable.

To explain the second requirement of \dnut, consider another term, $B \oplus N_A \oplus \penc{N_B}{\pk(A)} \oplus \senc{N_A}{N_B}$. We can introduce tags in this term as well, similar to the previous term, as $[1,B] \oplus [2,N_A] \oplus [3,\penc{N_B}{\pk(A)}] \oplus [4,\senc{N_A}{N_B}]$, so as to satisfy the first requirement. However, this would violate the second requirement, since interms in both might be unifiable. For instance, $[1,A]$ in the first term is $\std$-Unifiable with $[1,B]$ in the second. To avoid this, we can range the interms in the first $\eqop$ term from $1.1$ to $1.4$, and the second from $2.1$ to $2.4$. So the terms are now, 
$[1.1, A] \oplus [1.2, N_B] \oplus [1.3, \penc{N_A}{\pk(B)}] \oplus [1.4,\senc{N_B}{K}]$ and $[2.1,B] \oplus [2.2,N_A] \oplus [2.3,\penc{N_B}{\pk(A)}] \oplus [2.4,\senc{N_A}{N_B}]$. Obviously, they satisfy the third requirement as well.

Below we give a protocol that has multiple \txor{} terms, to illustrate how \dnut{} may be satisfied in protocols where there might be many complex and nested terms:

\begin{center}
\begin{tabular}{|ll|l|}
	\hline \hline 
\textbf{Original}	&	\textbf{protocol} 	&	\textbf{Changed to satisfy \dnut}  \\ 
	\hline \hline 
       $A \to B$ : & $A, B$ &    $A, B$ \\

      $B \to A$  : & $[N_B, B] \oplus \penc{N_B, A}{\pk(A)}$ &     $[2.1, N_B, B] \oplus [2.2, \penc{N_B, A}{\pk(A)}]$\\

			$A \to B$  : & $A \oplus N_B \oplus \penc{A \oplus N_B}{\pk(B)} \oplus $ &			  $[3.1, A] \oplus [3.2, N_B] \oplus [3.3, \penc{[3.3.1, A] \oplus $  \\

		& $\senc{N_A}{N_B}$  &     $[3.3.2, N_B]}{\pk(B)}] \oplus [3.4,\senc{N_A}{N_B}]$ \\   

		$A \to B$  : & $\penc{N_A \oplus N_B, A, B}{\pk(A)} \oplus$ &   $[4.1, \penc{[4.1.1, N_A] \oplus [4.1.2, N_B], A, B}{\pk(A)}] \oplus$ \\ 			
		&	$\senc{N_A \oplus A, N_B \oplus B}{N_A \oplus N_B}$ &			 $[4.2, \senc{[4.2.1, N_A] \oplus [4.2.2, A]$, $[4.3.1,N_B] \oplus$\\ 
		&		&    $[4.3.2, B]}{[4.4.1,N_A] \oplus [4.4.2, N_B]}]$  \\

	\hline
\end{tabular}
\end{center}

\section{Main Result}\label{s.main-result}

We will now prove that, if \dnut~is followed in a set of terms,  the effects of equational theories are totally disabled.

\begin{theorem}\label{t.main-result}

Let $T$ be a set of terms that are \dnutsat. Then, if two non-variables are unifiable in the $(\std \cup \eqth)$ theory, then they are also unifiable in the $(\std \cup \feqop)$ theory:

	$\text{\dnutsat}(T) \Ra (\fa m, t \in T)( (m, t \notin \Vars) \wedge (m \approx_{(\std \cup \eqth)} t) \Ra (m \approx_{(\std \cup \feqop)} t) )$.

\end{theorem}

\begin{proof}

Suppose $\{ \tuple{m}{t} \} = \Gamma$. 

From BSCA, for $m$ and $t$ to be $(\std \cup \eqth)$-Unifiable, every $\tuple{m_1}{t_1} \in  \Gamma_{5.1}$ should be $\std$-Unifiable and every $\tuple{m_1}{t_1} \in  \Gamma_{5.2}$ should be $\eqth$-Unifiable:

\begin{equation}\label{e.Gamma12}
	(\fa \tuple{m_1}{t_1} \in \Gamma_{5.1})(m_1 \approx_{\std} t_1) \wedge 	(\fa \tuple{m_1}{t_1} \in \Gamma_{5.2})(m_1 \approx_{\eqth} t_1).
\end{equation}

Suppose $\tuple{m_1}{t_1} \in \Gamma_{5.2}$. From BSCA, we have that, for $\tuple{m_1}{t_1}$ to be $\eqth$-Unifiable, for every interm $x$ of $m$, there should exist a term $y$ as an interm of $m_1$ or $t_1$ such that, $x$ and $y$ are $\std$-Unifiable (unless $x$ is the Unity element):

\begin{equation}\label{e.newvars-must-unify}
		(\fa x)
		\left(
			\begin{array}{c}
				 (x \Subset m_1) \wedge (x \neq 0) \Ra (\ex y)(((y  \Subset m_1) \vee (y \Subset t_1)) \wedge (x \approx_{\std} y)) 
			\end{array}
		\right).
\end{equation}

Now from \dnut{} Condition~3, no $\eqop$ term has the Unity element as an interm. From \dnut{} Condition~1, interms \emph{within} an $\eqop$ term should not be $\std$-Unifiable. Hence, $y$ cannot be an interm of $m_1$. Similarly, from \dnut{} Condition~2, interms \emph{between} two different $\eqop$ terms should not be $\std$-Unifiable as well. Hence, $y$ cannot be an interm of $t_1$ either. 

The only other way for $m_1$ and $t_1$ to be $\eqth$-Unifiable is that $m_1$ must be equal to $t_1$, in which case they are both $\eqth$-Unifiable and $\feqop$-Unifiable.

Thus in general, every $\tuple{m_1}{t_1}$ belonging to $\Gamma_{5.2}$ is $\feqop$-Unifiable. Further, from (\ref{e.Gamma12}), every $\tuple{m_1}{t_1}$ belonging to $\Gamma_{5.1}$ is $\std$-Unifiable.

Hence, $\tuple{m}{t}$ is $(\std \cup \feqop)$-Unifiable.

\end{proof}
\section{Conclusion}\label{s.concl}

In this paper, we showed that tagging messages that were constructed with operators possessing algebraic properties, disables the equational theories induced by those properties.

Tags specified in \dnut{} basically disable cancellation of terms entirely, both inside a term or between different terms. For {\sf ACUIdem} and {\sf ACUInverse}, no change is required at all in \dnut. For other theories that are disjoint with the standard theory, we can use similar tagging to disable cancellation, and disable the theories. In the presentation and the full paper, we will explain those details and also the impact of the main result on symbolic protocol analysis. 

The main result easily falls apart under homomorphic encryption (HE). For instance, the UP $[1,A] \stackrel{?}{\approx} [3,a] \oplus [6,b] \oplus [4,C]$ has \dnutsat{} terms. It is unifiable under $\std \cup \msf{HE}$ with $\{ a/A, b/C \}$ as the unifier, if binary encoding is used for the tags 3, 4 and 6, since $[3,a] \oplus [6,b] \oplus [4,C] = [011\oplus 110 \oplus 100, a \oplus b \oplus C]$ under HE, which is equal to $[1,a]$ if $C = b$. But it is not under $\std \cup \feqop$. 

It seems that extending the result under non-disjoint theories such as $\std$ and $\msf{HE}$ will be quite challenging. Although BSCA cannot be used, I conjecture that new unification algorithms such as~\cite{ALLNR09} might be useful in this pursuit. I look forward to discussions with the workshop participants toward further work in this direction.

\bibliographystyle{eptcs} 
\bibliography{UNIF10-eptcs}

\appendix
\section{Bader \& Schulz Combined theory unification algorithm}\label{s.BSCA}

In this section, we will describe Bader \& Schulz's combination algorithm \cite{BS96} (abbreviated to BSCA) that combines unification algorithms for two disjoint theories.

We will use the following $(\std \cup \acun)$-UP as our running example:

\[  \penc{1,n_a}{\mi{pk}(B)} \stackrel{?}{\approx_{\std \cup \acun}} \penc{1,N_B}{\mi{pk}(a)}   \oplus   [2,A] \oplus  [2,b]. \\  \]

\noindent
\paragraph*{Step 1 (Purify terms).} BSCA first ``purifies" the given set of $(Th = Th_1 \cup Th_2)$-unification problems, $\Gamma$, into a new set of problems $\Gamma_1$ through the introduction of some new variables such that, all the terms are ``pure" wrt $Th_1$ or $Th_2$, but not both.

If our running example was $\Gamma$, then, the set of problems in $\Gamma_1$ are $W \stackrel{?}{\approx_{\std}} [1,n_a]_{\mi{pk}(B)}$, $X  \stackrel{?}{{\approx_{\std}}}  [1,N_B]_{\mi{pk}(a)}, Y  \stackrel{?}{\approx_{\std}} [2,A]$, $Z  \stackrel{?}{\approx_{\std}}  [2,b]$, and $W \stackrel{?}{\approx_{\acun}} X \oplus Y \oplus Z$, where $W, X, Y, Z$ are obviously new variables that did not exist in $\Gamma$.

\noindent
\paragraph*{Step 2. (Purify problems)} Next, BSCA purifies the unification problems such that every problem in the set has both terms belonging to the same theory.
For our example problem, this step can be skipped since all the problems in $\Gamma_1$ already have both their terms purely from the same theory ($\std$ or $\acun$)). 

\paragraph*{Step 3. (Variable identification)} Next, BSCA partitions variables in $\Gamma_2$ into a partition $\mi{VarIdP}$ such that each variable in $\Gamma_2$ is replaced with a representative from the same equivalence class in $\VarIdP$. The result is $\Gamma_3$.

In our example problem, one set of values for $\VarIdP$ can be $\{\{A\},\{B\},\{N_B\}, \{W\},\{X\}$, \\  $\{Y,Z\}\}.$

\paragraph*{Step 4. (Split the problem)} The next step of BSCA is to split $\Gamma_3$ into two sets of problems such that each set $\Gamma_{4.i}$ has every problem with terms from the same theory, $Th_i$ ($i \in \{1,2\}$). 

Following this in our example, 

\[ \Gamma_{4.1} = \{ \tuple{W}{[1,n_a]_{\mi{pk}(B)}}, \tuple{X}{[1,N_B]_{\mi{pk}(a)}}, \tuple{Y}{[2,A]}, \tuple{Z}{[2,b]}  \}, \] 

and 

\[ \Gamma_{4.2} = \{ \tuple{W}{\xorseq{X}{Y}{Y}} \}.  \]

\paragraph*{Step 5. (Solve systems)} The penultimate step of BSCA is to partition all the variables in $\Gamma_3$ into a size of two: Let $p = \{ V_1, V_2 \}$ is a partition of $\Vars(\Gamma_3)$. Then, the earlier problems ($\Gamma_{4.1}$, $\Gamma_{4.2}$) are further split such that all the variables in one set of the partition are replaced with new constants in one of the set of problems and vice-versa in the other.

In our sample problem, we can form $\{ V_1, V_2 \}$ as $\{ \Vars(\Gamma_3), \{\} \}$. i.e., we choose that all the variables in problems of $\Gamma_{5.2}$ be replaced with new constants. This is required to find the unifier for the problem (this is the partition that will successfully find a unifier).

So $\Gamma_{5.1}$ stays the same as $\Gamma_{4.1}$, but $\Gamma_{5.2}$ is changed to $\Gamma_{5.2} = \Gamma_{4.2} \beta =	 \{ \tuple{W}{\xorseq{X}{Y}{Y}} \} \beta  =  \{ \tuple{w}{\xorseq{x}{y}{y}} \}$. i.e., $\beta = \{ w/W, x/X, y/Y \}$, where, $w, x, y$ are constants, which obviously did not appear in $\Gamma_{5.1}$. 

\paragraph*{Step 6. (Combine unifiers)} The final step of BSCA is to combine the unifiers obtained in Step 5 for $\Gamma_{5.1}$ and $\Gamma_{5.2}$:

\begin{definition}{\em\bf [Combined Unifier]}\label{d.Combined-Unifier}

Let $\Gamma$ be a $\Th$-UP where $(\Th_1 \cup \Th_2) = \Th$. Let $\sigma_i \in A_{\Th_i}(\Gamma_{5.i}$, $i \in \{1,2\}$ and let $V_i = \Vars(\Gamma_{5.i})$, $i \in \{1, 2 \}$.

Suppose `$<$' is a linear order on $\Vars(\Gamma)$ such that $Y < X$ if $X$ is not a subterm of an instantiation of $Y$:

	\[	(\fa X, Y \in \Vars(\Gamma))((Y < X) \Ra (\not\ex \sigma)(X \sqsubset Y\sigma)).  \]

Let $\msf{least}(X,T,<)$ be defined as the minimal element of set $T$, when ordered linearly by the relation `$<$'. i.e., 

\[	\msf{least}(X,T,<)	\Lra	(\fa Y \in T)((Y \neq X) \Ra (X < Y)).	\]

Then, the combined UA for $\Gamma$, namely $A_{\Th_1 \cup \Th_2}$, is defined such that,

\[	A_{\Th_1\cup \Th_2}(\Gamma) = \{ \sigma \mid (\ex \sigma_1,\sigma_2)((\sigma = \sigma_1 																					\odot \sigma_2) \wedge (\sigma_1 \in A_{\Th_1}(\Gamma_{5.1})) \wedge (\sigma_2 \in A_{\Th_2}(\Gamma_{5.2}))) \}.
\] 

\noindent
where, if $\sigma = \sigma_1 \odot \sigma_2$, then,

\begin{itemize}

\item The substitution in $\sigma$ for the least variable in $V_1$ and $V_2$ is  from $\sigma_1$ and $\sigma_2$ respectively: \\

$(\fa i \in \{ 1, 2 \})( (X \in V_i) \wedge \msf{least}(X, \Vars(\Gamma), <)  \Rightarrow (X\sigma = X\sigma_i))$; and \\

\item For all other variables $X$, where each $Y$ with $Y < X$ has a substitution already defined, define
$X\sigma = X\sigma_i\sigma$ $(i \in \{1,2\})$: \\

$(\fa i \in \{ 1, 2 \})( (\fa X \in V_i)( (\fa Y)( (Y < X) \wedge (\ex Z)(Z/Y \in \sigma) )) \Rightarrow (X\sigma = X \sigma_i \sigma))$.

\end{itemize}

\end{definition}

\end{document}